\documentclass[amsmath,amssymb]{revtex4}

\usepackage{graphicx}
\usepackage{dcolumn}
\usepackage{bm}
\usepackage{amsmath}
\usepackage{amssymb}

\newcommand{\be}{\begin{equation}}
\newcommand{\ee}{\end{equation}}

\begin{document}

\title{Free Energy of the Eight Vertex Model with an Odd Number of Lattice Sites}

\author{Ovidiu I. P\^{a}\c{t}u}

 \email{ipatu@grad.physics.sunysb.edu}
 \affiliation{C.N. Yang Institute for Theoretical Physics, State University of New York at Stony Brook, Stony Brook, NY 11794-3840, USA }
  \altaffiliation[Permanent address ]{Institute of Space Sciences MG 23, 077125 Bucharest-Magurele, Romania}

\date{\today}


\begin{abstract}

We calculate the bulk contribution for the doubly degenerated largest eigenvalue
of the transfer matrix of the eight vertex model with an  odd number of lattice
sites N in the disordered regime using the generic equation for roots
proposed by Fabricius and McCoy in \cite{Mccoy}.
We show as expected that in the thermodynamic limit
the result coincides with the one in the N even case.

\end{abstract}
\maketitle


\section{Introduction}

In the last years it has been seen that in  exactly solvable systems there are interesting features which  distinguish between systems with an an odd
and even number of lattice sites \cite{S1,S2,GBNM,TF0,BCR,DMN,PS,BSS,Sch,A}. The properties of finite size systems are qualitatively
different in the two cases and also totally new features occur in the odd case.

Recently in a series of papers \cite{Mccoy},\cite{Mccoy1}  Fabricius and McCoy have investigated the properties of the $\mathbf{Q}_{72}(v)$
matrix introduced by Baxter in Appendix C of \cite{Baxter2} in his study of the eight vertex model (for more information on the eight vertex model
see \cite{Baxter3}). This matrix is defined  for systems at
the root of unity

\be\label{ru}
\eta=\frac{mK}{L}
\ee
where $K$ is the complete elliptic integral of the first kind of modulus $k$, $m$ and $L$ are integers and the zeroes of the eigenvalues of $\mathbf{Q}_{72}(v)$ play an important role in the calculation of
the eigenvalues of the transfer matrix
of the eight vertex model.
The result of this investigation showed that the distribution of these zeroes in the case of an even number of lattice sites $N$
is qualitatively different from the $N$ odd case. In the $N$ even case these zeroes appear either as $L$-strings of the form
$u_k^L+2jK/L, j=0,\cdots,L-1$ or pairs $u_k^B$ and $u_k^B+iK'$
called "Bethe's roots" which satisfy "Bethe's equation"
\be\label{Betheequation}
\left(\frac{h(u_j^B+\eta)}{h(u_j^B-\eta)}\right)^N=
-e^{-2\pi i(\nu-n_B)m/L}\prod_{k=1}^{n_B}\frac{h(u_j^B-u_k^B+2\eta)}{h(u_j^B-u_k^B-2\eta)}
\ee
where $n_B$ is the number of pairs, $H(v)$ and $\Theta(v)$ are elliptic theta functions of modulus $k$
(for  definitions and some properties see Appendix \ref{definitions})
and $h(v)=H(v)\Theta(v)$.

In the $N$ odd case when $m$ is odd (this is the case where the matrix $\mathbf{Q}_{72}(v)$ exists \cite{Mccoy1,Mccoy})
we do not have Bethe roots or the $L$-strings. We do have however the following type of pairing: for every
set of roots $u_j$ we have the second pair of roots $u_j+iK'$ which also gives an eigenvalue of $\mathbf{Q}_{72}(v)$
and these two eigenvalues satisfy the "TQ" equation with the same eigenvalue of the transfer matrix $T(v)$.
These roots satisfy (3.3) of \cite{Mccoy}

\be\label{eq8}
\left(\frac{h(u_j+\eta)}{h(u_j-\eta)}\right)^N=
-e^{-\frac{2\pi i\nu\eta}{K}}\prod_{k=1}^N\frac{H(u_j-u_k+2\eta)}{H(u_j-u_k-2\eta)}
\ee
This equation differs from  "Bethe's equation"  (\ref{Betheequation}) in that the right hand side contains the
function $H(v)$ and not $h(v)$ and the number of terms in the product is $N$. For the $m$ even case numerical investigations
showed that the ground state has the same structure of roots and they obey also (\ref{eq8}).
The equation (\ref{eq8}) is new in the literature and the purpose of this paper is to calculate the largest
eigenvalue for the transfer matrix with $N$ odd  ($m$ even or odd) when the largest eigenvalue is doubly degenerate.

We make this observation more precise by focusing on the case with $N$ odd which is the main interest of this paper.
Baxter's \cite{Baxter2},\cite{Baxter1} original computation of the eigenvalues of the transfer matrix $\mathbf{T}(v)$ of the eight vertex
employs the so called "TQ" equation
\be\label{TQ}
\mathbf{T}(v)\mathbf{Q}_{72}(v)=[\rho\Theta(0)h(v+\eta)]^N\mathbf{Q}_{72}(v-2\eta)+[\rho\Theta(0)h(v-\eta)]^N\mathbf{Q}_{72}(v+2\eta)
\ee
where $\mathbf{Q}_{72}(v)$ is an auxiliary matrix with the properties
\be
[\mathbf{T}(v),\mathbf{Q}_{72}(v')]=[\mathbf{Q}_{72}(v),\mathbf{Q}_{72}(v')]=0
\ee
and $\eta$ is restricted at the root of unity condition (\ref{ru}).
Also in \cite{Baxter2} it has been showed that if we parametrize the vertex weights in terms of $\rho,v,\eta,k$ as
\be\label{weighta}
a(v)=\rho\Theta(2\eta)\Theta(v-\eta)H(v+\eta)
\ee
\be\label{weightb}
b(v)=\rho\Theta(2\eta)H(v-\eta)\Theta(v+\eta)
\ee
\be
c(v)=\rho H(2\eta)\Theta(v-\eta)\Theta(v+\eta)
\ee
\be
d(v)=\rho H(2\eta)H(v-\eta)H(v+\eta)
\ee
then transfer matrices with different values of the parameter $v$ commute
\be
[\mathbf{T}(v),\mathbf{T}(v')]=0
\ee
The operator $\mathbf{S}$ defined as
\be
\mathbf{S}=\prod_{j=1}^N\sigma^z
\ee
commutes with $\mathbf{T}(v)$ and $\mathbf{Q}_{72}(v)$
\be
[\mathbf{S},\mathbf{T}(v)]=[\mathbf{S},\mathbf{Q}_{72}(v)]=0
\ee
has eigenvalues $(-1)^{\nu'}=\pm 1$ and under spin inversion  $\sigma^z\rightarrow-\sigma^z $ we have $ \mathbf{S}\rightarrow(-1)^N \mathbf{S}$ .
Under spin inversion the eigenvalues of the transfer matrix are invariant so
in the case of $N$ odd the eigenvalues of the transfer matrix are doubly degenerate one with $\nu'=0$ and one with $\nu'=1$.
In the $N$ odd case the matrix $\mathbf{Q}_{72}(v)$ which satisfies
(1.5) and (1.6) of \cite{Mccoy} and the "TQ" equation exists only in the case when $\eta$
is of the form (\ref{ru}) with $m$ odd and
from \cite{Mccoy} we know that the eigenvalues of the $\mathbf{Q}_{72}$ matrix can be written in
the factorized form
\be
Q_{72}(v)=\mathcal{K}(q,u_k)e^{-\frac{i\pi\nu v}{2K}}\prod_{k=1}^N H(v-u_k)
\ee
with $\mathcal{K}(q,u_k)$ a constant and $u_k$ the zeroes of the eigenvalues of $\mathbf{Q}_{72}$.
As we have said in this case with $m$ odd and $L$ even or odd
the roots satisfy (\ref{eq8}) and we do not have Bethe roots or $L$-strings.


In the case of $m$ even with $L$ odd numerical investigations presented in \cite{Mccoy} revealed the following new features as $\eta\rightarrow mK/L$:
some of the roots moved to form $L$-strings giving a contribution which cancels from the generic equation of roots (\ref{eq8})
and the rest arranged in pairs $u_k$ and $u_k+iK'$. In this cases (\ref{eq8}) reduced to the "Bethe's equation" (\ref{Betheequation}).
However the roots of the solutions without the $L$-strings or paired roots still satisfy (\ref{eq8}). Also it should be mentioned
that in both cases ($m$ even or odd) the ground state is characterized by solutions with $N$ roots satisfying (\ref{eq8}).

These roots of the eigenvalues of the $\mathbf{Q}_{72}(v)$ are the main ingredient in calculating the eigenvalues of the transfer matrix.
The free energy per  site of the infinite lattice is related to the largest eigenvalue of the transfer matrix
by
\be
f=-k_BT\lim_{N\rightarrow\infty}N^{-1}\ln|T_{max}(N)|
\ee
where $k_B$ is the Boltzmann constant.

Originally in 1971  Baxter \cite{Baxter1} calculated the largest eigenvalues for the $N$ even case using a new perturbation-expansion technique.
In 1973 Johnson, Krinski and McCoy \cite{Mccoy2} reproduced the result of Baxter  and calculated also the next largest eigenvalues
using the integral-equation method which was applied previously to other models solvable by Bethe ansatz. In 1988 Kl\"umper and Zittartz {\cite{klu1}
using an inversion relation for the eigenvalues of the transfer matrix obtained all the eigenvalues and in 1989 they
classified the excited states \cite{klu2}.

In our paper for the calculation of the largest eigenvalue of the transfer matrix we are going to use the integral equation method
(Sect. IV of \cite{Mccoy2} and Sect. VI of \cite{TF}). The plan of the paper is as follows: in Section \ref{doi} we are going to obtain and solve the linear integral equation for the
density of roots. In Section \ref{trei} we are going to use the explicit expression for the density of roots to calculate the
largest eigenvalue of the transfer matrix in the thermodynamic limit (\ref{main1}) which is the main result of this paper. Using the result of Baxter
obtained for the N even case (for a sketch of the derivation in the disordered region see \cite{Bazhanov}) we will show that the bulk
contributions are the same. The technical details of the calculations are relegated to the Appendixes.
Our analysis will be restricted in the disordered regime; more precisely in the region

\be
0<q=e^{-\frac{\pi K'}{K}}<1
\ee
\be
0<\eta<K
\ee

with $K>0$ and $K'>0$ where $K$ is the complete elliptic integral of the first kind of modulus $k$
and $K'$ the same integral of complementary modulus.
Also we are employing the convention that the theta-functions whose modulus is not explicit should be understood as $k$.


\section{Integral Equation for the Density of Roots for N Odd}\label{doi}

The main feature of the eight vertex model with $N$ odd is the fact that the largest eigenvalue of the
transfer matrix is doubly degenerate. In order to calculate it (for both $m$ even and odd)
we are going to use (\ref{eq8}) the generic equation of roots proposed by Fabricius and McCoy in \cite{Mccoy} and
the following set of roots (see Table I in \cite{Mccoy})
\be\label{set1}
  u_j^1=iv_j+K \mbox{ with } -K'<v_j<K' \ \
\ee
\be\label{set2}
  u_j^2=u_j^1+iK'                       \ \
\ee
Using the sum rules obtained in  \cite{Mccoy}
\be
N+\sum_{j=1}^N\frac{Re(u_j)}{K}=\mbox{ even integer }
\ee
\be
\nu=\sum_{j=1}^N\frac{Im(u_j)}{K'}=\mbox{ even integer}-\nu'-N
\ee
we have for the first set of roots $\{u_j^1\}, \nu=0,\nu'=1$ and for the second set of roots $\{u_j^2\},\nu=N,\nu'=0$.

Using (\ref{tetat1}) and (\ref{tetat}) eq. \ref{eq8} for the largest eigenvalue can be written for both sets of roots as
\be\label{ger}
\left(\frac{H_1(iv_j+\eta)\Theta_1(iv_j+\eta)}{H_1(-iv_j+\eta)\Theta_1(-iv_j+\eta)}\right)^N=
\prod_{k=1}^N\frac{H(iv_j-iv_k+2\eta)}{H(-iv_j+iv_k+2\eta)}\ \ \ \ (j=1,\cdots,N)
\ee
Define the functions
\be
A_p(x,\xi)=i\ln\frac{H_1(p\xi-ix)\Theta_1(p\xi-ix)}{H_1(p\xi+ix)\Theta_1(p\xi+ix)},\ \ \xi\in \mathbb{R}
\ee
\be
B_p(x,\xi)=i\ln\frac{H(p\xi-ix)}{H(p\xi+ix)},\ \ \xi\in \mathbb{R}
\ee
Both of these functions are analytic in a strip around the real axis (see Appendix \ref{AB}) defined by the conditions
\be
|\mathcal{I} mx|+|p\xi |<K
\ee
and
\be
|\mathcal{I} mx|+|(p\xi-K) |<K
\ee
respectively. In terms of these functions (\ref{ger}) takes the form
\be
NA_1(v_j,\eta)=2\pi I_j+\sum_{k=1}^{N}B_2(v_j-v_k,\eta)\ \ \ \ (j=1,\cdots,N)
\ee
where $I_j$ are integers specifying the branches of the logarithm. For the largest eigenvalue we have $I_j-I_{j-1}=-1, j=2,\cdots,N$.
As $N\rightarrow\infty$ the number of roots becomes infinite and their imaginary parts fill the interval $[-K',K']$. The number of roots
with the imaginary parts in the
interval $[x,x+dx]$ is given by $N\rho_O(x)dx$ where $\rho_O(x)$ is the density of roots.
Using
\be
\frac{1}{N}\sum_{k=1}^Nf(v_k)=\int_{-K'}^{K'}f(x)\rho_O(x)dx,\ \ N\rightarrow\infty
\ee
in the large $N$ limit the coupled nonlinear equations are replaced by the linear integral equation
\be\label{lie}
A'_1(x,\eta)=-2\pi\rho_O(x)+\int_{-K'}^{K'}B'_2(x-x')\rho_O(x')dx'
\ee
where the prime denotes the derivative  with respect to $x$. This equation can be solved using the Fourier transform (see Appendix \ref{rootskernel})
obtaining
\be
\rho_O(x)=\frac{1}{2K'}\sum_{n=-\infty}^\infty\widehat{\rho}_O(n)e^{-\frac{in\pi x}{K'}}
\ee
with
\begin{eqnarray}
\widehat{\rho}_O(0)&=&1\nonumber\\
\widehat{\rho}_O(n=\mbox{odd})&=&0\\
\widehat{\rho}_O(n=\mbox{even})&=&\frac{1}{\cosh\left[n\left(\frac{\pi K}{K'}-\frac{\pi \eta}{K'}\right)\right]}\nonumber
\end{eqnarray}


\section{Largest Eigenvalue of the Transfer Matrix}\label{trei}

It can be shown (see Appendix \ref{relevant} for a proof in the domain $v+2\eta<K,0<v,\eta$) that in the "TQ" equation
\begin{eqnarray}
T_{NO}(v,\nu')&=&[\rho\Theta(0)h(v+\eta)]^N \frac{Q_{72}(v-2\eta,\nu')}{Q_{72}(v,\nu')}+[\rho\Theta(0)h(v-\eta)]^N \frac{Q_{72}(v+2\eta,\nu')}{Q_{72}(v,\nu')}\nonumber\\
         &=&T_{NO}^+(v,\nu')+T_{NO}^-(v,\nu')
\end{eqnarray}
$T_{NO}^+(v,\nu')$ exponentially dominates $T_{NO}^-(v,\nu')$  for large $N$ so we will focus our attention on the first term.
From \cite{Mccoy} we know that the eigenvalues of the $\mathbf{Q}_{72}$ matrix can be written in
the factorized form
\be\label{deqe}
Q_{72}(v,\nu')=\mathcal{K}(q,u_k)e^{-\frac{i\pi\nu v}{2K}}\prod_{k=1}^N H(v-u_k)
\ee
with $\mathcal{K}(q,u_k)$ a constant and $u_k$ the zeroes of the eigenvalues of $\mathbf{Q}_{72}$.
Explicitly we have for $T_{NO}^+(v,\nu')$ the following expressions for the first set of roots (\ref{set1}) when $\nu=0, \nu'=1$
\be
T_{NO}^+(v,\nu'=1)=[\rho\Theta(0)h(v+\eta)]^N \prod_{k=1}^{N}\frac{H(iv_k+K-v+2\eta)}{H(iv_k+K-v)}
\ee
and for the second set of roots (\ref{set2}) when $\nu=N, \nu'=0$, using (\ref{tetat1})
\be
T_{NO}^+(v,\nu'=0)=[\rho\Theta(0)h(v+\eta)]^N \prod_{k=1}^{N}\frac{\Theta(iv_k+K-v+2\eta)}{\Theta(iv_k+K-v)}
\ee
After we have used (\ref{tetat}) and the Jacobi transformation (\ref{jh1}) and (\ref{jt1})
we obtain
\be\label{lep}
T_{NO}^+(v,\nu')=[\rho\Theta(0)h(v+\eta)]^N \prod_{k=1}^{N}e^{\frac{\pi(\eta v-\eta^2-i\eta v_k)}{KK'}}
                         \frac{\Theta(v_k+iv-2i\eta+(1-\nu')K',k')}{\Theta(v_k+iv+(1-\nu')K',k')}
\ee

Defining
\be
g_{O}^+(x,\nu')=\ln\left(e^{\frac{\pi(\eta v-\eta^2-i\eta x)}{KK'}}
                         \frac{\Theta(x+iv-2i\eta+(1-\nu')K',k')}{\Theta(x+iv+(1-\nu')K',k')}\right)
\ee
and taking the logarithm of (\ref{lep}) we have
\be\label{intermediate}
\ln T_{NO}^+(v,\nu')=N \ln[\rho\Theta(0)h(v+\eta)] +\sum_{k=1}^{N}  g_{O}^+(v_k,\nu')
\ee
As in the previous section the sum in (\ref{intermediate}) can be transformed in
\be\label{le}
\sum_{k=1}^{N}  g_{O}^+(v_k,\nu')=N\int_{-K'}^{K'}dxg_{O}^+(x,\nu')\rho_O(x)
\ee
or in terms of Fourier coefficients
\begin{eqnarray}\label{sumtoint}
\sum_{k=1}^{N}g_{O}^+(v_k,\nu')&=&\frac{N}{2K'}\sum_{n=-\infty}^{+\infty}\widehat{g}_{O}^+(n,\nu')\widehat{\rho}_O(-n)
\end{eqnarray}
It follows from (\ref{texp}) and (\ref{t1exp}) that
\be
g_{O}^+(x,\nu')=\frac{\pi(\eta v-\eta^2-i\eta x) }{KK'}+\sum_{n=1}^\infty\frac{2(-1)^{[(1-\nu')n]}}{n\sinh\left(\frac{n\pi K}{K'}\right)}
                            \sin\left(\frac{n\pi (x+iv-i\eta)}{K'}\right)\sin\left(-\frac{n\pi i\eta}{K'}\right)
\ee
valid for
\be
|\mathcal{I}m(x+iv-i\eta)|+\eta<K
\ee
so the Fourier coefficients are
\begin{eqnarray}
\widehat{g}_{O}^+(0,\nu')&=&\frac{2\pi}{K}(\eta v -\eta^2)\nonumber\\
\widehat{g}_{O}^+(n,\nu')&=&\frac{2K'}{n}\left( e^{\frac{n\pi(v-\eta)}{K'}}\frac{(-1)^{[(1-\nu')n]}\sinh\left(\frac{n\pi \eta}{K'}\right)}{\sinh\left(\frac{n\pi K}{K'}\right)}
                     -\frac{\eta}{ K}\cos(n\pi)\right)
\end{eqnarray}


Collecting all the results and noting the fact the Fourier coefficients of the density of roots
are zero for $n$ odd  we obtain for the largest eigenvalue of the transfer matrix (for both sets of roots)

\begin{eqnarray}\label{main1}
\frac{1}{N}\ln T_{NO}^+(v,\nu')&=&\ln[\rho\Theta(0)h(v+\eta)]+\frac{\pi}{KK'}(\eta v-\eta^2)\nonumber\\
         &+&\sum_{n=1}^{+\infty}
         \frac{\sinh\left[2n\left(\frac{\pi v}{K'}-\frac{\pi\eta}{K'}\right)\right]\sinh\left(\frac{2n\pi \eta}{K'}\right)}{n\sinh\left(\frac{2n\pi K}{K'}\right)\cosh\left[2n\left(\frac{\pi K}{K'}-\frac{\pi \eta}{K'}\right)\right]}\nonumber\\
\end{eqnarray}
in the domain
\be\label{dom}
0<\eta<v<K
\ee
Even if our result was obtained in the domain (\ref{dom}) we see that we can consider the analytical continuation of (\ref{main1})
which is valid in
\be
0<\eta<K,\ \ \  \eta<v<2K-\eta
\ee
Now we will show that  our result for the largest eigenvalue of the transfer matrix in the disorder regime (\ref{main1}) coincides
with the result  obtained for even $N$ (A.51 of \cite{Bazhanov}). The $N$ even result obtained for the normalization
factor $\rho'=2H_1(0|q^{1/2})^{-1}\Theta(0|q)^{-1}$ is
\be\label{BM}
\frac{1}{N}\ln T_{NE}^+(v')=\ln\theta_1(v'+\eta'|q')+\frac{2i\eta'(v'-\eta')}{\pi\tau}+\sum_{n=1}^\infty\frac{\sinh\left(\frac{2in\eta'}{\tau}\right)\sinh\left(\frac{2in(v'-\eta')}{\tau}\right)}                                                    {n\sinh\left(\frac{in\pi}{\tau}\right)\cosh\left(\frac{in(\pi-2\eta')}{\tau}\right)}
\ee
with
\be
\eta'=\frac{\pi\eta}{2K}\ \ \ \ v'=\frac{\pi v}{2K}\ \ \ \ \tau=\frac{iK'}{2K}\ \ \ \ q'=e^{i\pi\tau}\ \ \ \ H(x)=\theta_1\left(\frac{\pi x}{2K}|q'^2\right)
\ee
It can be easily seen that (\ref{BM}) and  (\ref{main1}) are equal if we consider in our result the same normalization factor
and the identity (\ref{Landen}).

\par

\section{Discussion and Conclusions}

We have calculated the doubly degenerated largest eigenvalue of the transfer matrix of the eight vertex model with
an odd number of lattice sites in the thermodynamic limit.
The next  interesting problem would be the calculation of the low lying  excited  states for the odd case in a manner
similar with the computations in \cite{Mccoy2}.  In the $m$ odd case the calculations
should be accessible, however in the $m$ even and $L$ odd case an interesting problem occurs. As we have said in the $m$ even and $L$ odd case
some of the roots move to form $L$-strings which drop from the generic equation for roots (\ref{eq8}) while the rest arrange
themselves in pairs. For these states the usual procedure \cite{Mccoy2} of calculating the energy of the excitations using linear integral equations
cannot be applied. This means that in this case the excitation spectrum has two different kind of states. The $m$ even and $L$ odd  case is different
 because in this case some of the eigenvalues develop extra degeneracies beyond the double degeneracy
present in the odd case. The $\eta=2K/3$ case is particularly interesting due to the fact that in this case it is believed that
the degenerated largest eigenvalue is exact even for a finite lattice
\be
\frac{1}{N}\ln T_{NO}(v)=a+b,\ \ \ \eta=2K/3
\ee
where $a$ and $b$ are the Boltzman weights of the model (see formulae \ref{weighta} and \ref{weightb}).
This was advocated in \cite{Baxter5,S3,BM} and verified analytically for $N$ up to 15 in \cite{Bazhanov}.

\acknowledgements
The author would like to thank Prof. Barry M. McCoy
for helpful  suggestions and comments.This work is partially supported by the National Science Foundation
under grant DMR-0302758.


\appendix

\section{Elliptic Theta Functions}\label{definitions}

Let $K$ be the complete elliptic integral of the first kind of modulus $k$ , $K'$ the
same integral of the complementary modulus $k'=(1-k^2)^{1/2}$ and $q=e^{-\pi K'/K}$.
The Jacobi theta-functions of nome $q$ are defined by the relations

\be
H(x)=2q^{1/4}\sin\left(\frac{\pi x}{2K}\right)\prod_{n=1}^\infty\left(1-2q^{2n}\cos\left(\frac{\pi x}{K}\right)+q^{4n}\right)(1-q^{2n})
\ee
\be
H_1(x)=2q^{1/4}\cos\left(\frac{\pi x}{2K}\right)\prod_{n=1}^\infty\left(1+2q^{2n}\cos\left(\frac{\pi x}{K}\right)+q^{4n}\right)(1-q^{2n})
\ee
\be
\Theta(x)=\prod_{n=1}^\infty\left(1-2q^{2n-1}\cos\left(\frac{\pi x}{K}\right)+q^{4n-2}\right)(1-q^{2n})
\ee\be
\Theta_1(x)=\prod_{n=1}^\infty\left(1+2q^{2n-1}\cos\left(\frac{\pi x}{K}\right)+q^{4n-2}\right)(1-q^{2n})
\ee

They obey the following relations

\be\label{tetat}
H(x+K)=H_1(x),\ \ \ \
\Theta(x+K)=\Theta_1(x)
\ee

\be
H(x+2K)=-H(x),\ \ \ \
\Theta(x+2K)=\Theta(x)
\ee

\be\label{tetat1}
\Theta(x\pm iK')=\pm iq^{-1/4}e^{\mp\frac{i\pi x}{2K}}H(x),\ \ \ \ H(x\pm iK')=\pm iq^{-1/4}e^{\mp\frac{i\pi x}{2K}}\Theta(x)
\ee

Using the fact that

\be
\Theta(x)=\prod_{n=1}^\infty(1-q^{2n-1}e^{-\frac{i\pi x}{K}})(1-q^{2n-1}e^{\frac{i\pi x}{K}})(1-q^{2n})
\ee

and

\be
H(x)=2q^{1/4}\sin\left(\frac{\pi x}{2K}\right)\prod_{n=1}^\infty(1-q^{2n}e^{-\frac{i\pi x}{K}})(1-q^{2n}e^{\frac{i\pi x}{K}})(1-q^{2n})
\ee

we can obtain the following identity (which is just the Landen transform)

\be\label{Landen}
H(x|q)\Theta(x|q)=H(x|q^{1/2})H_1(0|q^{1/2})/2
\ee

where we have explicitly expressed the dependence on the nome.

Also we have the following Fourier expansions

\be\label{texp}
\ln \frac{\Theta(x+y)}{\Theta(x-y)}=\sum_{n=1}^\infty\frac{2}{n\sinh\left(\frac{n\pi K'}{K}\right)}
                                   \sin\left(\frac{n\pi x}{K}\right)\sin\left(\frac{n\pi y}{K}\right)
\ee

valid for $|\mathcal{I} mx|+|\mathcal{I} my|<K'$

\be\label{t1exp}
\ln \frac{\Theta_1(x+y)}{\Theta_1(x-y)}=\ln \frac{\Theta(x+y-K)}{\Theta(x-y+K)}=
                                   \sum_{n=1}^\infty\frac{2(-1)^n}{n\sinh\left(\frac{n\pi K'}{K}\right)}
                                   \sin\left(\frac{n\pi x}{K}\right)\sin\left(\frac{n\pi y}{K}\right)
\ee

valid for $|\mathcal{I} mx|+|\mathcal{I} my|<K'$

\begin{eqnarray}\label{Hexp}
\ln \frac{H(x+y)}{H(x-y)}&=&\ln\left( -e^{-\frac{i\pi x}{K}}\frac{\Theta(x+y-iK')}{\Theta(x-y+iK')}\right)\nonumber\\
                         &=&i\left(-\pi-\frac{\pi x}{K}\right)+\sum_{n=1}^\infty\frac{2}{n\sinh\left(\frac{n\pi K'}{K}\right)}
                            \sin\left(\frac{n\pi x}{K}\right)\sin\left(\frac{n\pi(y-iK')}{K}\right)
\end{eqnarray}

valid for $|\mathcal{I} mx|+|\mathcal{I} m(y-iK')|<K'$

\begin{eqnarray}\label{H1exp}
\ln \frac{H_1(x+y)}{H_1(x-y)}&=&\ln\left( -\frac{H(x+y-K)}{H(x-y+K)}\right)\nonumber\\
                         &=&-\frac{i\pi x}{K}+\sum_{n=1}^\infty\frac{2(-1)^n}{n\sinh\left(\frac{n\pi K'}{K}\right)}
                            \sin\left(\frac{n\pi x}{K}\right)\sin\left(\frac{n\pi(y-iK')}{K}\right)
\end{eqnarray}

valid for $|\mathcal{I} mx|+|\mathcal{I} m(y-iK') |<K'$



\section{ Analysis of the relevant functions}\label{AB}

In order to facilitate our analysis  of $ A_p(x,\xi)$ and $B_p(x,\xi)$ functions we are going to perform  Jacobi transforms
on the theta-functions involved. The relevant formulae are (Chap. XV of \cite {Baxter3} or Chap. XXI of \cite{WW})

\be
H(ix,k)=i\sqrt{K'/K} e^{\frac{\pi x^2}{4KK'}}H(x,k')
\ee

\be\label{jh1}
H_1(ix,k)=\sqrt{K'/K} e^{\frac{\pi x^2}{4KK'}}\Theta(x,k')
\ee

\be
\Theta(ix,k)=\sqrt{K'/K} e^{\frac{\pi x^2}{4KK'}}H_1(x,k')
\ee

\be\label{jt1}
\Theta_1(ix,k)=\sqrt{K'/K} e^{\frac{\pi x^2}{4KK'}}\Theta_1(x,k')
\ee

where we have explicitly expressed the modulus dependence of the theta-functions involved.

\par
\par
After the Jacobi transformation and using (\ref{texp}) and (\ref{t1exp}) we obtain for $A_p$

\begin{eqnarray}\label{2exp}
A_p(x,\xi)&=&i\ln\left(\frac{H_1(p\xi-ix,k)\Theta_1(p\xi-ix,k)}{H_1(p\xi+ix,k)\Theta_1(p\xi+ix,k)}\right)\nonumber\\
               &=&i\ln\left(e^{\frac{2i\pi p\xi x}{KK'}}\frac{\Theta(ip\xi+x,k')\Theta_1(ip\xi+x,k')}{\Theta(-ip\xi+x,k')\Theta_1(-ip\xi+x,k')}\right)\nonumber\\
               &=&-\frac{2\pi p\xi x}{KK'}+\sum_{n=1}^\infty\frac{2i}{n\sinh\left(\frac{2n\pi K}{K'}\right)}
                            \sin\left(\frac{2n\pi x}{K'}\right)\sin\left(\frac{2n\pi ip\xi}{K'}\right)
\end{eqnarray}

where the expansion is valid for $|\mathcal{I} mx|+|\mathcal{I}m (ip\xi)|<K$ and we specify the branches of the logarithms
by also demanding that no cuts of $A_p$ cross   the real $x$ axis and $A_p(0,\xi)=0$.

Analogously after the Jacobi transformation and using (\ref{Hexp}) we obtain for $B_p$

\begin{eqnarray}\label{3exp}
B_p(x,\xi)&=&i\ln\left(\frac{H(p\xi-ix,k)}{H(p\xi+ix,k)}\right)\nonumber\\
               &=&i\ln\left((-)e^{\frac{i\pi p\xi x}{KK'}}\frac{H(ip\xi+x,k')}{H(-ip\xi+x,k'))}\right)\nonumber\\
               &=&\left(1-\frac{p\xi}{K}\right)\frac{\pi x}{K'}+\sum_{n=1}^\infty\frac{2i}{n\sinh\left(\frac{n\pi K}{K'}\right)}
                            \sin\left(\frac{n\pi x}{K'}\right)\sin\left(\frac{n\pi(ip\xi-iK)}{K'}\right)
\end{eqnarray}

where the expansion is valid for $|\mathcal{I} mx|+|\mathcal{I} m(ip\xi-iK) |<K$ and we specify the branches of the logarithms
by demanding that no cuts of $B_p$ cross   the real $x$ axis and $B_p(0,\xi)=0$.



\section{Density of Roots }\label{rootskernel}

We define the Fourier coefficients of a function periodic (quasiperiodic) of period (quasiperiod) $2K'$
by
\be
\widehat{f}(n)=\int_{-K'}^{K'}dx f(x)e^{\frac{in\pi x}{K'}}
\ee
In terms of this coefficients the original function is expressed as
\be
f(x)=\frac{1}{2K'}\sum_{n=-\infty}^\infty \widehat{f}(n)e^{-\frac{in\pi x}{K'}}
\ee
After the Fourier transform the linear integral equation for the roots takes the form
\be
\widehat{A}'_1(n,\eta)=-2\pi\widehat{\rho}_O(n)+\widehat{B}'_2(n,\eta)\widehat{\rho}_O(n)
\ee
so
\begin{eqnarray}
\rho_O(x)&=&\frac{1}{2K'}\sum_{n=-\infty}^{+\infty}\frac{-\widehat {A}'_1(n,\eta)}{2\pi-\widehat {B}_2'(n,\eta)}
       e^{-\frac{in\pi x}{K'}}\nonumber\\
\end{eqnarray}

In order to obtain explicit expressions for the density of roots we need to calculate
the Fourier coefficients of $B_2'$ and $A_1'$. From (\ref{3exp}) we obtain the following expansion
\be
B_2(x,\eta)=\left(1-\frac{2\eta}{K}\right)\frac{\pi x}{K'}+\sum_{n=1}^\infty\frac{2i}{n\sinh\left(\frac{n\pi K}{K'}\right)}
                            \sin\left(\frac{n\pi x}{K'}\right)\sin\left(\frac{n\pi(2i\eta-iK)}{K'}\right)
\ee
so the Fourier coefficients for $B_2'$ are
\begin{eqnarray}\label{f30}
\widehat{B}_2'(0,\eta)&=&2\pi\left(1-\frac{2\eta}{K}\right)\nonumber\\
\widehat{B}_2'(n,\eta)&=&2\pi\frac{\sinh\left[n\left(\frac{\pi K}{K'}-\frac{2\pi\eta}{K'}\right)\right]}
                      {\sinh\left(\frac{n\pi K}{K'}\right)}
\end{eqnarray}
In the case of $A_1'$  using  (\ref{2exp}) we have the following expansion
\be
A_1(x,\eta)=-\frac{2\pi \eta x}{KK'}+\sum_{n=1}^\infty\frac{2i}{n\sinh\left(\frac{2n\pi K}{K'}\right)}
                            \sin\left(\frac{2n\pi x}{K'}\right)\sin\left(\frac{2n\pi i\eta}{K'}\right)
\ee
so the Fourier coefficients are
\begin{eqnarray}\label{f20}
\widehat{A}_1'(0,\eta)&=&-\frac{4\pi\eta}{K}\nonumber\\
\widehat{A}_1'(n=\mbox{odd},\eta)&=&0\\
\widehat{A}_1'(n=\mbox{even},\eta)&=&-4\pi\frac{\sinh\left(\frac{n\pi\eta}{K'}\right)}{\sinh\left(\frac{n\pi K}{K'}\right)}\nonumber
\end{eqnarray}

Using (\ref{f30}) and (\ref{f20}) and the identity
\be
\sinh a+\sinh(2b-a)=2\sinh b\cosh(b-a)
\ee
we obtain
\begin{eqnarray}
\widehat{\rho}_O(0)&=&1\nonumber\\
\widehat{\rho}_O(n=\mbox{odd})&=&0\\
\widehat{\rho}_O(n=\mbox{even})&=&\frac{1}{\cosh\left[n\left(\frac{\pi K}{K'}-\frac{\pi \eta}{K'}\right)\right]}\nonumber
\end{eqnarray}



\section{Relevant Term in the "TQ" Equation}\label{relevant}

The logarithm of the ratio of the two terms in the "TQ" equation is

\be\label{ratio}
\ln\frac{T_{NO}^+(v)}{T_{NO}^-(v)}=N\ln\left(\frac{h(v+\eta)}{h(v-\eta)}\right)+\ln\left(\frac{Q_{72O}(v-2\eta)}{Q_{72O}(v+2\eta)}\right)
\ee

Let's analyze the first term in the RHS of (\ref{ratio}). After the Jacobi transformations it becomes

\be
N\ln\left(\frac{h(v+\eta)}{h(v-\eta)}\right)=N\ln\left( e^{-\frac{2\pi v\eta}{KK'}}\frac{H(iv+i\eta,k')H_1(iv+i\eta,k')}{H(iv-i\eta,k')H_1(iv-i\eta,k')}\right)
\ee

so using (\ref{Hexp}) and (\ref{H1exp}) we obtain

\be\label{firstterm}
N\ln\left(\frac{h(v+\eta)}{h(v-\eta)}\right)=-N\frac{2\pi v\eta}{KK'}+iN\left(-2\pi-\frac{2\pi iv}{K}\right)+\sum_{n=1}^\infty\frac{2N}{n\sinh\left(\frac{2n\pi K}{K'}\right)}
                            \sinh\left(\frac{2n\pi \eta}{K'}\right)\sinh\left(\frac{2n\pi(K-v)}{K'}\right)
\ee
where the sum is a positive number ($0<\eta<v<K$).

For the second term in the RHS of (\ref{ratio}) using (\ref{deqe}),(\ref{tetat}) and the Jacobi transformation we obtain

\begin{eqnarray}
\ln\left(\frac{Q_{72O}(v-2\eta)}{Q_{72O}(v+2\eta)}\right)&=&\sum_{i=1}^N\ln\left(\frac{H_1(iv_j-v+2\eta)}{H_1(iv_j-v-2\eta)}\right)\\
                      &=&\sum_{i=1}^N\ln\left(e^{-\frac{2\pi i\eta(v_j+iv)}{KK'}}\frac{\Theta(v_j+iv-2i\eta,k')}{\Theta(v_j+iv+2i\eta,k')}\right)
\end{eqnarray}

Defining
\be
g(x)=\ln\left(e^{-\frac{2\pi i\eta(x+iv)}{KK'}}\frac{\Theta(x+iv-2i\eta,k')}{\Theta(x+iv+2i\eta,k')}\right)
\ee

the sum can  be written as
\be
\sum_{k=1}^{N}g(v_j)=N\left(\frac{1}{2K'}\sum_{n=-\infty}^{+\infty}\widehat{g}(n)\widehat{\rho}_O(-n)+\mathcal{O}(\frac{1}{N})\right)
\ee

From (\ref{texp}) we have the following expansion
\be
g(x)=\frac{2\pi v\eta }{KK'}-\frac{2i\pi\eta x}{KK'}+\sum_{n=1}^\infty\frac{2}{n\sinh\left(\frac{n\pi K}{K'}\right)}\sin\left(\frac{n\pi(x+iv)}{K'}\right)
                                                      \sin\left(-\frac{2in\pi\eta}{K'}\right)
\ee

so we obtain

\be\label{secondterm}
\sum_{k=1}^{N}g(v_j)=N\left( \frac{2\pi v \eta}{KK'}+\sum_{n=-\infty,n\ne 0 }^\infty\frac{e^{\frac{2n\pi v}{K'}}\sinh\left(\frac{4n\pi\eta}{K'}\right)}
                          {2n\sinh\left(\frac{2n\pi K}{K'}\right)\cosh\left[2n\left(\frac{\pi K}{K'}-\frac{\pi \eta}{K'}\right)\right]}+\mathcal{O}(\frac{1}{N})\right)
\ee

From (\ref{firstterm}) and (\ref{secondterm}) we see that

\be
\ln\frac{T_{NO}^+(v)}{T_{NO}^-(v)}=Nc+\mathcal{O}(\frac{1}{N})
\ee

where $c$ is a number whose real part is positive.

\end{document}